\documentclass[aps,prl,twocolumn,groupedaddress,nofootinbib,showpacs]{revtex4}
\usepackage[dvips]{graphicx}
\newif\ifdraft
\drafttrue
\newif\ifpreprint

\preprintfalse

\def\fig#1{fig.~{\ref{#1}}}

\def\eqn#1{eq.~({\ref{#1}})}
\def\nn{\nonumber}
\def\NeqFour{\mathcal{N}=4}
\def\NeqEight{\mathcal{N}=8}

\def\eps{\epsilon}

\def\tree{{\rm tree}}

\def\Tr{{\rm Tr}}

\def\f{\tilde f}

\newbox\charbox
\newbox\slabox
\def\s#1{{      
        \setbox\charbox=\hbox{$#1$}
        \setbox\slabox=\hbox{$/$}
        \dimen\charbox=\ht\slabox
        \advance\dimen\charbox by -\dp\slabox
        \advance\dimen\charbox by -\ht\charbox
        \advance\dimen\charbox by \dp\charbox
        \divide\dimen\charbox by 2
        \raise-\dimen\charbox\hbox to \wd\charbox{\hss/\hss}
        \llap{$#1$}
}}

\begin{document}

\title{
\ifpreprint
\hbox{\rm \small
UCLA/12/TEP/105 $\null\hskip 4.7cm\null$ \hfill 
SU-ITP-12/22$\null\hskip 4.7cm\null$ \hfill 
Saclay--IPhT--T12/056 \break}
\hbox{$\null$ \break}
\fi
The Five-Loop Four-Point Amplitude of $\NeqFour$ Super-Yang-Mills Theory
}

\author{Z.~Bern${}^{a}$, J.~J.~M.~Carrasco${}^{b}$,
 H.~Johansson${}^{c}$ and R.~Roiban${}^{d}$}

\affiliation{
${}^a$\hbox{Department of Physics and Astronomy, UCLA, Los Angeles, 
CA 90095, USA}\\
${}^b$\hbox{Stanford Institute for Theoretical Physics and 
Department of Physics,}\\ 
\hbox{Stanford University, Stanford, CA 94305, USA}\\
${}^c$\hbox{Institut de Physique Th\'eorique, CEA--Saclay,  F--91191
Gif-sur-Yvette cedex, France}\\
${}^d$\hbox{Department of Physics, Pennsylvania State University,
University Park, PA 16802, USA}
\\
}

\begin{abstract}
Using the method of maximal cuts, we construct the complete
$D$-dimensional integrand of the five-loop four-point amplitude of
$\NeqFour$ super-Yang-Mills theory, including nonplanar
contributions. In the critical dimension where this amplitude becomes
ultraviolet divergent, we present a compact explicit expression for
the nonvanishing ultraviolet divergence in terms of three vacuum
integrals.  This construction provides a crucial step towards
obtaining the corresponding amplitude of $\NeqEight$ supergravity useful
for resolving the general ultraviolet behavior of supergravity theories.
\end{abstract}

\pacs{04.65.+e, 11.15.Bt, 11.30.Pb, 11.55.Bq \hspace{1cm}}

\maketitle


Recent years have seen remarkable progress in understanding and
constructing scattering amplitudes in gauge and gravity theories,
driven largely by the advent of on-shell techniques.  The advances
have had broad applications including computations in quantum
chromodynamics of multijet processes at the Large Hadron Collider,
resummations of planar $\NeqFour$ super-Yang-Mills (sYM) amplitudes
linking them to string theory via the AdS/CFT correspondence,
connections to integrability of planar $\NeqFour$ sYM theory and
studies of ultraviolet (UV) and infrared diverges in gauge and gravity
theories. (See ref.~\cite{AmplitudeReviews} for recent reviews.)

These advances have been most striking for the maximally
supersymmetric $\NeqFour$ sYM amplitude, in the planar limit where the
number of color charges is large. Significant progress has also been
made for the less well understood nonplanar case, which is the subject
of this Letter.  Nonplanar contributions to amplitudes in $\NeqFour$
sYM theory have been obtained previously through four
loops~\cite{BRY,BDDPR, CompactThree, JJHenrikFivePt, Neq44np, ck4l},
along with detailed studies of their UV properties in higher
space-time dimensions.  The planar part of the five-loop amplitude is
found in ref.~\cite{FiveLoop}.  Here we carry out a similar study for
the five-loop four-point amplitude and analyze the UV divergences in
the lowest dimension where they occur. We anticipate that our results
will become useful for detailed studies of the structure of the
theory, including infrared singularities, anomalous dimensions and
other observables related to amplitudes. Such studies can be a useful
laboratory for quantum chromodynamics, for example, to help resolve
the full structure of infrared singularities (see
e.g.\ ref. \cite{QCDIR}).

Beyond the intrinsic interest for understanding $\NeqFour$ sYM theory,
our construction of the five-loop four-point amplitude is a key step
towards obtaining the corresponding amplitudes of ${\cal N} \ge 4$
supergravity, needed to help resolve the long-standing question on the
possible UV finiteness these theories.  In fact, whenever a
representation of an $\NeqFour$ sYM amplitude is constructed that
exhibits a duality between color and kinematics~\cite{BCJ, BCJLoop},
a simple pathway exists for obtaining corresponding ${\cal N} \ge
4$ supergravity loop integrands~\cite{JJHenrikFivePt, N4Grav12Loop,
  N4Grav, ck4l}. In particular, the $\NeqEight$ supergravity integrand
follows trivially when the duality is manifest, simply by
replacing color factors with the kinematic numerators of the
diagrams.  Although the form of the five-loop four-point amplitude
presented here does not manifest the required duality, it does offer
an excellent starting point for finding such representations.

Explicit constructions of amplitudes have played a key role for
determining the UV divergence structure of gauge and gravity theories
as a function of dimension. $\NeqFour$ sYM theory is
known~\cite{BDDPR,HoweStelleRevisited} to be UV finite in
dimensions
\begin{equation}
D < 4 + \frac{6}{L} \,,  \hskip 1 cm   (L>1)
\label{FinitenessBound}
\end{equation}
where $L$ is the loop order. This exhibits the well-known UV
finiteness in $D=4$~\cite{Mandelstam}.  A remaining open question is
whether the bound (\ref{FinitenessBound}) is saturated to all loop
orders.  From explicit computations, it is known to be saturated for
$L\le 4$~\cite{BDDPR,CompactThree,Neq44np,ck4l}.  As commented on in
ref.~\cite{Neq44np}, the $L=5$ planar amplitude~\cite{FiveLoop} also
saturates the bound (\ref{FinitenessBound}).  Below we give a simple
expression for the divergence, including nonplanar parts.

A related open question is whether maximally supersymmetric
$\NeqEight$ supergravity has the same finiteness bound as $\NeqFour$
sYM theory, implying it is UV finite in $D=4$, or if it has a worse
behavior.  (For a recent optimistic opinion see ref.~\cite{Kallosh};
for a recent pessimistic one see ref.~\cite{Banks}.)  Explicit calculations
of the divergences~\cite{BDDPR,GravityThree, GravityFour,
  CompactThree, ck4l} and symmetry and other
arguments~\cite{SevenLoopGravity} show that through four loops, the
bound (\ref{FinitenessBound}) holds in $\NeqEight$ supergravity.
Beyond this, the arguments suggest that $\NeqEight$ supergravity will
have a worse behavior, leading to a seven-loop divergence in $D=4$.
However, when similar symmetry arguments are applied to $\NeqFour$
supergravity, they imply the existence of a valid three-loop
counterterm~\cite{VanishingVolume}; the coefficient of this
counterterm has recently been explicitly shown to
vanish~\cite{N4Grav}. (See ref.~\cite{PierreN4} for a string-based
argument.) This exhibits better behavior than implied by known
symmetry considerations and is in line with cancellations suggested by
unitarity arguments~\cite{Finite}.  In particular, it emphasizes the
importance of directly checking the amplitudes whether
\eqn{FinitenessBound} holds for $\NeqEight$ supergravity at $L=5$.


\begin{figure}[t]
\includegraphics[clip,scale=0.5]{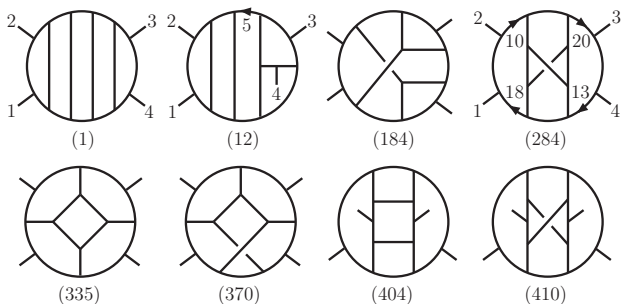}
\caption[a]{Sample graphs for the five-loop four-point $\NeqFour$ sYM
  amplitude. The graph labels correspond to the ones used in the  
  ancillary file~\cite{AttachedFile}. }
\label{SampleGraphsFigure}
\end{figure}

Our construction of the five-loop four-point amplitude of $\NeqFour$
sYM theory organizes it in the form,
\begin{equation}
\hskip -.01 cm 
{\cal A}_4^{(5)} = i g^{12} st A_4^\tree \sum_{{\cal S}_4} \sum_{i =1}^{416}
\! \int \!  \prod_{j=5}^9\frac{d^D l_j} {(2\pi)^D} \frac{1}{S_i}
 \frac{C_i\, N_i}{\prod_{m=5}^{20} l_{m}^2} \,, 
\label{AmplitudeGraphs}
\end{equation}
where the second sum runs over a set of 416 distinct (non-isomorphic)
graphs with only cubic (trivalent) vertices. Some sample graphs are
shown in \fig{SampleGraphsFigure}.  The first sum runs over all 24
permutations of external leg labels indicated by ${\cal S}_4$.  The
symmetry factors $S_i$ remove overcounts, including those arising from
internal automorphism symmetries with external legs fixed.  Here we
absorb all contact terms (i.e. terms with fewer than the maximum
number of propagators) into graphs with only cubic vertices, by
multiplying and dividing by appropriate propagators. We denote
external momenta by $k_i$ for $i=1,\ldots,4$ and the five independent
loop momenta by $l_j$ for $j=5,\ldots,9$.  The remaining $l_j$ are
linear combinations of these.  The color factors $C_i$ of all graphs
are obtained by dressing every three-vertex in the graph with a factor
of $\f^{abc} =\Tr([T^{a},T^{b}]T^{c})$, where the gauge group
generators $T^a$ are normalized as $\Tr(T^a T^b) = \delta^{ab}$.  The
gauge coupling is $g$ and the crossing symmetric prefactor $st
A_4^\tree$ is in terms of the color-ordered $D$-dimensional tree
amplitude $A_4^\tree \equiv A_4^\tree(1,2,3,4)$ and $s= (k_1 + k_2)^2$
and $t =(k_2 + k_3)^2$.

\begin{figure}[t]
\includegraphics[clip,scale=0.5]{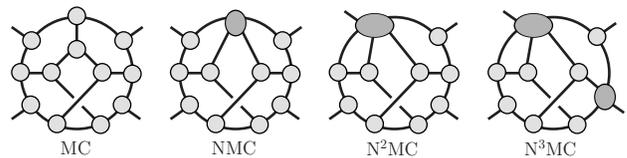}
\caption[a]{Sample N$^k$-maximal cuts for $k=0,1,2,3$.  The exposed 
lines are all cut.}
\label{MaxCutsFigure}
\end{figure}

\begin{figure}[t]
\includegraphics[clip,scale=0.36]{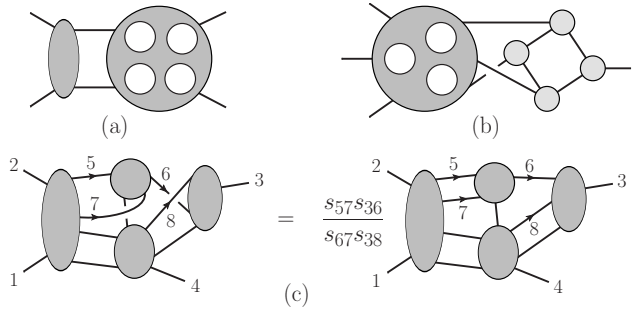}
\caption[a]{Examples of simple cuts used to speed up the calculation.
  (a) is a two particle cut, (b) a box cut and (c) is a sample
  application of the new amplitude relations of ref.~\cite{BCJ}.  The
  exposed lines are all cut.}
\label{SimpleCutsFigure}
\end{figure}

To construct the numerators $N_i$, we use the method of maximal
cuts~\cite{FiveLoop}, based on the unitarity
method~\cite{UnitarityMethod}.  Application of this method and various
strategies for greatly streamlining the construction of the numerators
has been described in considerable detail in ref.~\cite{Neq44np}, so
here we give only a brief summary. The method works in $D$ dimensions
and can be used to obtain local expressions, from which UV divergences
can be straightforwardly extracted.

We start with an ansatz for the diagram numerators
containing free parameters to be determined by matching against
generalized unitarity cuts.  Our ansatz is a polynomial of degree four
in the kinematic invariants, subject to the power-counting constraint
that no term has more than six powers of loop momentum.  We also
demand that each numerator respects the automorphism symmetries of the
graph.  Once a solution is found satisfying a complete set of cut
conditions, we have the integrand.  If an inconsistency is
encountered, the ansatz must be enlarged.  We note that the solutions for
numerators are not unique and different choices can be mapped into each
other by generalized gauge transformations~\cite{BCJ,BCJLoop,BCJSquare}.

The parameters of the ansatz are determined from generalized unitarity
cuts that decompose a loop integrand into products of on-shell tree
amplitudes summed over all intermediate states, $\sum_{\rm states}
A^\tree_{(1)} A^\tree_{(2)} \cdots A^\tree_{(m)}$. These cuts are
organized according to the number of cut propagators that are replaced
with on-shell conditions.  We start from the maximal cuts (MCs) where
all 16 internal propagators cut.  After obtaining the MCs, we then
constructs all next-to-maximal cuts (NMCs), with 15 cut propagators.
We continue this process, systematically constructing analytic
expressions for (next-to)$^k$-maximal cuts (N$^k$MCs) with fewer and
fewer imposed cut conditions. For the five-loop four-point $\NeqFour$
sYM amplitude this process terminates at $k=3$, since the power
counting of the theory prevents numerator factors from canceling more
than 3 propagators.  Representative cuts for $k=0,1,2,3$ are shown in
\fig{MaxCutsFigure}. The number of nonzero (color-stripped) cuts of
type N$^k$MC are 410, 2473, 7917, 15156 for $k=0,1,2,3$,
respectively. This count does not include the different independent
color orderings of each cut.  In addition to the nonvanishing cuts,
there is a large class of N$^{k\le3}$MCs that evaluate to zero because
they contain nontrivial ($n\le3$)-point subamplitudes.

Each cut can be reduced to a relatively simple analytic expression.
All N$^k$MCs used in the construction are evaluated in $D$ dimensions
by embedding them in auxiliary cuts that can be directly expressed in
terms of simplified analytic forms.  As discussed in some detail in
ref.~\cite{Neq44np}, two particularly useful cuts for this purpose are
two-particle cuts and box cuts. Whenever a two-particle reducible cut
can be factorized into two four-point amplitudes, as illustrated in
\fig{SimpleCutsFigure}(a), all contributions to the cut can be written
down immediately using lower-loop results~\cite{BRY}.  Similarly, all
cut contributions that possess a four-point loop or box subdiagram,
illustrated in \fig{SimpleCutsFigure}(b), are simple to
evaluate~\cite{Neq44np}.  A third type of auxiliary cut,
illustrated in \fig{SimpleCutsFigure}(c), allows us to map known
$D$-dimensional planar cuts to nonplanar ones via the new
tree-amplitude relations uncovered in ref.~\cite{BCJ}.  Alternatively,
one can construct numerator representations that obey the
color-kinematics duality for each cut separately~\cite{Neq44np},
giving local numerator relations between planar and nonplanar
diagrams, up to terms that vanish on the cut. This technique is
especially useful whenever the cut contains massless bubble or tadpole
subdiagrams (as sometimes occurs for N$^3$MCs), since the local
numerators are automatically free of spurious singularities that can
appear with other methods.  A fourth type of auxiliary cut valid in
$D$ dimensions and used in our construction is one that maps five-loop
nonplanar cuts to simpler six-loop planar
cuts~\cite{CarrascoJohanssonNonplanar}.

We have found a choice of parameters in the starting ansatz whose cuts
correctly reproduce the N$^k$MCs at the level of the integrand. We
thus have a complete integral representation of the five-loop
amplitude.  As a few simple examples, the numerators of graphs 1, 12
and 284 are
\begin{eqnarray}
N_{1} &=& s^4\,, \hskip 1 cm N_{12} =2s^3 k_3\cdot l_5\,, \nn \\
N_{284} &=& 2s^2 ((l_{10}\cdot l_{20})^2+(l_{13} \cdot l_{18})^2 )\,,
\end{eqnarray}
corresponding to the graphs in \fig{SampleGraphsFigure} labeled as
(1), (12) and (284) and matching the labeling in the ancillary
file~\cite{AttachedFile}. The lines with arrows in
\fig{SampleGraphsFigure} give the momentum labels and directions. The
symmetry factors $1/S_i$ for these graphs are, respectively, $1/4$, $1$
and $1/4$.

The complete set of 416 nonvanishing graphs with their associated
symmetry factors, numerators and color factors are included in the
ancillary file~\cite{AttachedFile}.  We note that graphs 61, 67, 133,
137, 263, 382, 412 have vanishing color factors for a general gauge
group (due to symmetry properties of the graph), and hence do not
contribute to the amplitude.  However, we include them in our
representation because they are needed for constructing gravity
amplitudes~\cite{Neq44np}.

We have carried out extensive cross checks on our result.  The cut
construction automatically cross checks the vast majority of
contributions because they are detected in multiple independent
channels. As an additional rather nontrivial check, in four dimensions
we confirmed numerically that all the analytically-obtained cuts are
correct; to carry out this check we used the simple algorithms of
ref.~\cite{SuperSum} for carrying out the supersums appearing in the
cuts.  We have also carried out systematic cross checks using
generalized cuts with up to six collapsed propagators. Furthermore, we
have evaluated a set of cuts that suffices to detect all ``snail''
contributions, equivalent to bubbles on external legs (see sections 2D
and 3C of ref.~\cite{ck4l}), showing that such contributions do not
appear in our representation.


\begin{figure}[t]
\includegraphics[clip,scale=0.5]{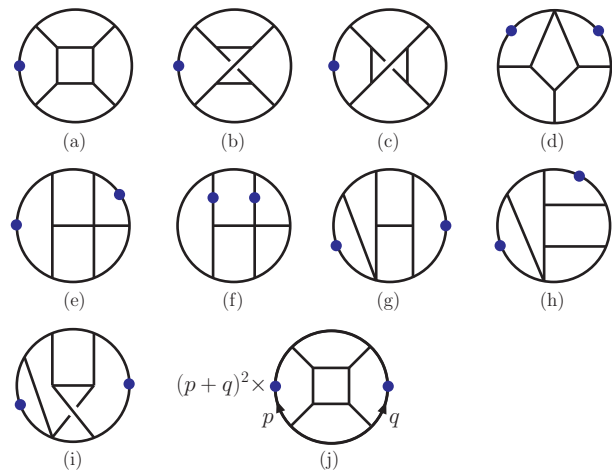}
\caption[a]{Some of the five-loop vacuum integrals that appear in
  intermediate steps.  Only (a), (b) and (c) appear in the final UV
  divergence. Integral (j) has a nontrivial numerator factor, as
  indicated. The (blue) dots indicate that a propagator is squared.}
\label{VacuumsFigure}
\end{figure}

Starting with the constructed integrand, we obtain the potential
logarithmic divergence in the five-loop critical dimension, $D=26/5$,
following the same strategy as at lower loops~\cite{GravityThree,
  GravityFour, Neq44np, ck4l}: We expand the amplitudes at small
external momentum and keep the leading term.  The result of this
expansion is a sum of about 185 vacuum diagrams; a few of which are
displayed in \fig{VacuumsFigure}.  As discussed in
refs.~\cite{GravityFour, ck4l}, the vacuum integrals in this expansion
are not all independent (so the precise number appearing initially can
vary).  We derive consistency relations between the vacuum integrals
by considering auxiliary linearly divergent integrals of similar
propagator structure, expanding them around zero external momenta and
requiring that the results of the expansion be independent of
different integrand parametrizations.  This also directly cross checks
the procedure for integral reduction since we obtain a highly
overconstrained set of homogeneous consistency equations. The fact
that no positive definite integral is set to zero by this system is
a strong check on the calculation. These consistency relations
eliminate most of the vacuum diagrams.  Two examples are,
\begin{eqnarray}
V^{\rm (j)}&=& \frac{24}{5} V^{\rm (a)} -2 V^{\rm (d)}\,, \nn \\
V^{\rm (b)} &=&  2 V^{\rm (c)} + 35 V^{\rm (i)}+ \frac{365}{6} V^{\rm (d)}
    - \frac{4175}{162} V^{\rm (e)} \nn \\
&& \null
    - \frac{1045}{18} V^{\rm (f)} - \frac{9865}{81} V^{\rm (g)} 
     +\frac{305}{3} V^{\rm (h)} \,, \hskip 1 cm 
\end{eqnarray}
where the labels correspond to the ones in \fig{VacuumsFigure}.

After using the consistency relations, the leading UV divergence is
remarkably simple and given by only three vacuum integrals.  For
$SU(N_c)$, it is 
\begin{eqnarray}
{\cal A}_4^{(5)}\Bigr|_{\rm div} \hskip -.2 cm &=&
 \! \frac{144}{5}  g^{12} s t A_4^\tree
N_c^3 \, \Bigl( N_c^2 V^{\rm (a)} \nn \\
&& \null
\hskip 1 cm   + 12 (V^{\rm (a)} + 2 V^{\rm (b)} + V^{\rm (c)}) \Bigr) \nn \\
&& \null \times
( t \f^{a_1 a_2 b} \f^{b a_3 a_4} + s \f^{a_2 a_3 b} \f^{b a_4 a_1} ) \,. \hskip 1 cm 
\label{FinalVacuum}
\end{eqnarray}
With the chosen normalization, the  Wick-rotated vacuum
integrals in \eqn{FinalVacuum} are all positive definite, proving that
no further hidden cancellations remain at $L=5$ in the critical
dimension for either leading- or subleading-color contributions.
Using {\sc FIESTA}~\cite{FIESTA} we have numerically evaluated the
integrals giving,
\begin{eqnarray}
V^{\rm (a)} = \frac{0.331 K}{\eps} \,, \hskip .23 cm 
V^{\rm (b)} = \frac{0.310 K}{\eps} \,, \hskip .23 cm 
V^{\rm (c)} = \frac{0.291 K}{\eps}   \,, \nonumber \\
\end{eqnarray}
where the dimensional regularization parameter is $\eps \equiv
(26/5-D)/2$, $K = 1/(4 \pi)^{13}$ and numerical integration
uncertainties are below the displayed digits.  It is interesting that
the ratio between the subleading and leading contributions
$45.0/N_c^2$ is rather close to the three- and four-loop ratios,
$43.3/N_c^2$ and $44.4/N_c^2$~\cite{Neq44np}.  A striking feature of
the result (\ref{FinalVacuum}) is that the divergence does not contain
terms beyond ${\cal O}(1/N_c^2)$ suppression, nor does it contain
double-trace contributions when converted to an $SU(N_c)$ color-trace
representation, in line with expectations from lower
loops~\cite{Neq44np}.  The second of these features has already been
discussed in
refs.~\cite{Neq44np,DoubleTraceNonrenormalization}. Furthermore, the
three integrals and their relative coefficients have a remarkable
similarity with the corresponding ones at four loops, as seen
by comparing to eq.~(5.33) of ref.~\cite{Neq44np}. At lower loops,
exactly the same combination of integrals appearing in the
subleading-color contributions to the $\NeqFour$ sYM divergences
appears in the corresponding ones of $\NeqEight$
supergravity~\cite{Neq44np}.  A natural conjecture is that the same
holds at five loops, so that the two theories share the same critical
dimension, $D=26/5$.


In summary, the five-loop amplitude we have constructed here offers
detailed information on the structure of the nonplanar sector of
$\NeqFour$ sYM theory. As a first application, we have shown that
simple patterns for divergences in the dimension where they first
appear continue to hold through five loops; this hints that the
divergences are controlled by a deep structure of the theory.  Our
construction of the five-loop four-point amplitude is an excellent
starting point to try to find a representation exhibiting the duality
between color and kinematics.  We expect that the results presented
here will be crucial input for obtaining corresponding supergravity
amplitudes and for studying their UV behavior.

\vskip .2 cm 

We thank S.~Davies, T.~L.~Dennen, S.~Ferrara, Y.-t.~Huang, R.~Kallosh,
D.~A.~Kosower, V.~A.~Smirnov, K.~Stelle and A.~Tseytlin for helpful
discussions. We especially thank L.J.~Dixon for collaboration on
related topics and for many helpful discussions and encouragement.
This research was supported by the US Department of Energy under
contracts DE--FG03--91ER40662 and DE--FG02--90ER40577 (OJI), and by
the US National Science Foundation under grant PHY-0756174. H.~J.'s
research is supported by the European Research Council under Advanced
Investigator Grant ERC-AdG-228301.

\end{document}